\documentclass[11pt,twoside]{article}
\usepackage{epsfig}
\usepackage{here}
\usepackage{amsmath}

\newcommand{\bm}{\bibitem}
\newcommand{\tauiso}{{\mbox{\boldmath $\tau$}}}
\newcommand{\gmu}{{\gamma_\mu}}

\begin{document}

\begin{center}{\bf\Large  Dynamics of strangeness production  
in the near threshold nucleon-nucleon collisions{\footnote{Talk
presented in the second symposium on threshold meson production
in $pp$ and $pd$ interactions, Jagellonian University, Cracow, Poland,
May 31-June 3, 2004.}}}
\end{center}
\vspace{0.5cm}
\begin{center}
   Radhey Shyam$^{a}$ \\[0.1cm]
   {\small \em 
         $^a$Saha Institute of Nuclear Physics, Calcutta, India and\\
         Department of Radiation Sciences, Uppsala University, Uppsala,
  Sweden }
\end{center}
\vspace{0.5cm}
\begin{center}
 \parbox{0.9\textwidth}{
  \small{
    {\bf Abstract:}\
We investigate the associated strangeness production reactions
$pp \rightarrow p\Lambda K^+$ and $pp \rightarrow p\Sigma^0 K^+$ within
an effective Lagrangian model. The initial interaction between the
two nucleons is modeled by the exchange of $\pi$, $\rho$, $\omega$,
and $\sigma$ mesons and the strangeness production proceeds via 
excitations of  $N^*(1650)$, $N^*(1710)$, and $N^*(1720)$ baryonic
resonance states. The parameters of the model at the nucleon-nucleon-meson
vertices are determined by fitting the  elastic nucleon-nucleon scattering
with an effective interaction based on the exchange of these four mesons,
while those at the resonance vertices are calculated from the known
decay widths of the resonances and from the vector meson dominance
model. Experimental data taken recently by the COSY-11 collaboration are 
described well by this approach. The one-pion-exchange diagram dominates
the production process at both higher and lower beam energies. 
The excitation of the $N^*$(1650) resonance contributes predominantly to 
both the production channels at near threshold energies. Our model with final
state interaction effects among the outgoing particles included within
the Watson-Migdal approximation, is able to explain the observed beam
energy dependence of the ratio of the total cross sections of these two
reactions.  } 
 }
\end{center}

\vspace{0.5cm}
\section{Introduction}

In recent years, there has been a considerable amount of interest
in the study of the associated strangeness production reactions
in proton-proton ($pp$) collisions. This is expected to provide
information on the manifestation of quantum chromodynamics (QCD)
in the non-perturbative regime of energies larger than those of
the low energy pion physics where the low energy theorem and partial
conservation of axial current (PCAC) constraints provide a useful
insight into the relevant physics~\cite{eric88}. The strangeness
quantum number introduced by this reaction leads to new degrees of
freedom into this domain which are expected to probe the admixture
of $\bar{s}s$ quark pairs in the nucleon wave function~\cite{albe96}
and also the hyperon-nucleon and
hyperon-strange meson interactions~\cite{delo89,adel90}.

The elementary nucleon-nucleon-strange meson production cross sections
are the most important ingredients in the transport model studies of 
the $K^+$-meson production in the nucleus-nucleus collisions, which
provide information on not only the initial collision dynamics but
also the nuclear equation of state at high
density~\cite{mosel91,brown91,maru94,misk94,hart94,liko94,liko95,liko98}.
Furthermore, the enhancement in the strangeness production 
has been proposed as a signature for the formation of the
quark-gluon plasma in high energy nucleus-nucleus
collisions~\cite{rafe82,knol88}.

The measurements performed in late 1960s and 1970s provided the data on
the total cross sections for the associated hyperon ($Y$)-kaon production
at beam momenta larger than 2.80 GeV/c (these cross sections are
listed in Ref.~\cite{land88}).
With the advent of the high-duty proton-synchrotron (COSY) at the
Forschungszentrum, J\"ulich, it has become possible to perform
systematic studies of the associated strangeness production at beam
momenta very close to the reaction threshold (see, e.g.,Ref.~\cite{mos02}
for a comprehensive review). At the near threshold beam energies, the
final state interaction (FSI) effects among the outgoing particles are
significant. Therefore, the new set of data are expected to probe also
the hyperon-nucleon and hyperon-strange meson interactions.

A very interesting result of the studies performed by the COSY-11
collaboration is that the ratio ($R$) of the total cross sections for
the $pp \to p\Lambda K^+$ and $pp \to p\Sigma^0 K^+$ reactions
(to be referred as $\Lambda K^+$ and $\Sigma^0 K^+$ reactions,
respectively) at the same excess energy
(defined as $\epsilon = \sqrt{s}-m_p-m_Y-m_K$, with $m_p$, $m_Y$, and
$m_K$ being the masses of proton, hyperon, and kaon respectively and $s$
the invariant mass of the collision), is about $28^{+6}_{-9}$ for
$\epsilon$ $<$ 13 MeV~\cite{sew99}. This result is very
intriguing because at higher beam energies ($\epsilon \approx$ 1000 MeV) 
this ratio is only around 2.5.

Several calculations have been reported \cite{gas00,sib00,lag01}
to explain this result. Assuming that the $\pi$- and $K$- exchange
processes are the only mechanism leading to the strangeness production,
the authors of Ref.~\cite{gas00} show within a (non-relativistic)
distorted wave Born approximation (DWBA) model that while the
$\Lambda K^+$ reaction is dominated by the $K$-exchange only, both
$K$- and $\pi$-exchange processes play an important role in the case of
$\Sigma^0 K^+$ reaction.Therefore, if the amplitudes
corresponding to the two exchanges in the latter case interfere
destructively, the production of $\Sigma^0$ is suppressed as compared
to that of $\Lambda$. It should however, be noted that in Ref.~\cite{sib00},
$K$- and $\pi$- exchange amplitudes are reported to be of similar
magnitudes for both $\Lambda K^+$ and $\Sigma^0K^+$ reactions.
 
Nevertheless, a conclusive
evidence in support of the relative signs of $\pi$- and $K$- exchange
amplitudes being opposite to each other is still lacking. Furthermore,
other mechanisms like excitation, propagation, and decay of intermediate
baryonic resonances which play (see, e.g., \cite{shy99,shy01,col97}) an
important role in the strangeness production, have not been considered 
by these authors. In the calculations reported in Ref.~\cite{lag01} also
the relative sign of $K-$ and $\pi-$ exchange terms is chosen solely 
by the criteria of reproducing the experimental data, although in this
work the theory has been applied to describe a wider range of data 
(which includes the polarization transfer results of the DISTO
experiment~\cite{bal99} and the missing mass distribution obtained in
the inclusive $K^+$ production measurements performed at
SATURNE~\cite{sie94} apart from the ratio $R$).
 
We have investigated the $\Lambda K^+$ and $\Sigma^0 K^+$ reactions at
near threshold as well as higher beam energies
in the framework of an effective Lagrangian approach (ELA)
\cite{shy99,shy01,shy96,shy98}. In this theory, the
initial interaction between two incoming nucleons is
modeled by an effective Lagrangian which is based on the exchange
of the $\pi$-, $\rho$-, $\omega$-, and $\sigma$- mesons. The coupling
constants at the nucleon-nucleon-meson vertices are determined by
directly fitting the T-matrices of the nucleon-nucleon ($NN$) scattering
in the relevant energy region. The ELA uses the pseudovector (PV) coupling
for the nucleon-nucleon-pion vertex which is consistent with the chiral
symmetry requirement of the quantum chromodynamics~\cite{wei68}. In
contrast to some earlier calculations~\cite{sib99}, both ($\Lambda K^+$
and $\Sigma^0K^+$) reactions proceed via excitation of the $N^*$(1650),
$N^*$(1710), and $N^*$(1720) intermediate baryonic resonance states.
The interference terms between the amplitudes of various resonances
are retained. To describe the near threshold data, the FSI effects in
the final channel are included within the framework of the Watson-Migdal
theory~\cite{wat52,shy98}. ELA has been used to describe
rather successfully the $pp \to pp\pi^0$, $pp \to pn\pi^+$
\cite{shy98,shy96}, $pp \to p K^+Y$ \cite{shy99,shy01} as well
as $pp \to ppe^+e^-$~\cite{shy03} reactions.
 
\section{Description of the Model}

We consider the tree-level structure (Fig.~1) of the amplitudes for 
the associated $K^+ Y$ production in proton-proton collisions, which   
proceeds via the excitation of the $N^*$(1650)($\frac{1}{2}^{-}$),
$N^*$(1710)($\frac{1}{2}^{+}$), and $N^*$(1720)($\frac{3}{2}^{+}$) 
intermediate resonances. To evaluate these amplitudes within the
effective Lagrangian approach, one needs to know the effective Lagrangians
(and the coupling constants appearing therein) at (a) the
nucleon-nucleon-meson, (b) the resonance-nucleon-meson, and (c) the
resonance-$K^+$-hyperon vertices.
\begin{figure}[H]
\parbox{0.5\textwidth}
   {\epsfig{file=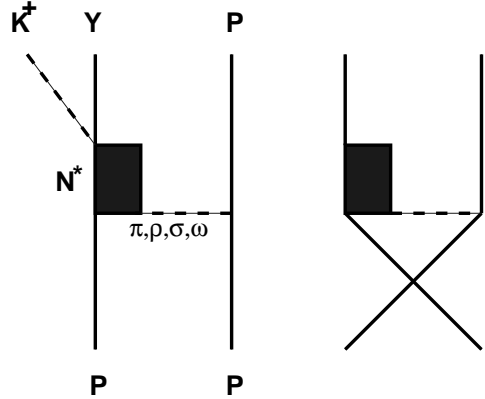,width=0.5\textwidth}}
\hfill
\parbox{0.4\textwidth}
  {\caption{\label{fig.1} {\small  
 Feynman diagrams for $K^{+}Y$ production in $pp$
collisions. The diagram on the left shows the direct process while that
on the right the exchange one.  }}}
\end{figure}
\noindent
The parameters for $NN$ vertices are determined by fitting the $NN$
elastic scattering T matrix with an effective $NN$ interaction based
on the $\pi$, $\rho$, $\omega$ and $\sigma$ meson exchanges. The
effective meson-$NN$ Lagrangians are  
\begin{eqnarray}
{\cal L}_{NN\pi} & = & -\frac{g_{NN\pi}}{2m_N} {\bar{\Psi}}_N \gamma _5
                             {\gamma}_{\mu} \tauiso
                            \cdot (\partial ^\mu {\bf \Phi}_\pi) \Psi _N. \\
{\cal L}_{NN\rho} &=&- g_{NN\rho} \bar{\Psi}_N \left( \gmu + \frac{k_\rho}
                         {2 m_N} \sigma_{\mu\nu} \partial^\nu\right)
                          \tauiso \cdot \mbox{\boldmath $\rho$}^\mu \Psi_N. \\
{\cal L}_{NN\omega} &=&- g_{NN\omega} \bar{\Psi}_N \left( \gmu + \frac{k_\omega}
                         {2 m_N} \sigma_{\mu\nu} \partial^\nu\right)
                          \omega^\mu \Psi_N.   \\
{\cal L}_{NN\sigma} &=& g_{NN\sigma} \bar{\Psi}_N \sigma \Psi_N.
\end{eqnarray}
We have used the notations and conventions of Bjorken and Drell~\cite{bjor64}. 
In Eq.~(1) $m_N$ denotes the nucleon mass.  Note that we have used
a PV coupling for the $NN\pi$ vertex.
Since we use these Lagrangians to directly model the T-matrix, we have
also included a nucleon-nucleon-axial-vector-isovector vertex, with the
effective Lagrangian given by
\begin{eqnarray}
{\cal L}_{NNA} & = & g_{NNA} {\bar {\Psi}} \gamma_5 \gamma_\mu \tauiso \Psi
                     \cdot {\bf {A}}^\mu,
\end{eqnarray}
where $A$ represents the axial-vector meson field. This term is introduced
because in the limit of large axial meson mass ($m_A$) it cures the 
unphysical behavior in the angular distribution of $NN$ scattering caused by
the contact term in the one-pion exchange amplitude~\cite{sch94}, if
$g_{NNA}$ is chosen to be
\begin{eqnarray}
g_{NNA} =  \frac{1}{\sqrt{3}} m_A \left(\frac{f_\pi}{m_\pi}\right),
\end{eqnarray}
with very large ($\gg m_N$) $m_A$. $f_\pi$ appearing in Eq.~(6) is 
related to $g_{NN\pi}$ as $f_\pi = (\frac{g_{NN\pi}}{2m_N})m_\pi$.

We introduce, at each interaction vertex, the form factor
\begin{eqnarray}
F_{i}^{NN} & = & \left (\frac{\lambda_i^{2} - m_i^{2}}{\lambda_i^{2} - q_i^{2}}
        \right ), i= \pi, \rho, \sigma, \omega,
\end{eqnarray}
where $q_i$ and $m_i$ are the four momentum and mass of the $i$th 
exchanged meson, respectively. The form factors suppress the contributions of
high momenta and the parameter $\lambda_i$, which governs the
range of suppression, can be related to the hadron size. Since
$NN$ elastic scattering cross sections decrease gradually with the beam
energy (beyond certain value), we take energy dependent meson-nucleon
coupling constants of the following form 
\begin{eqnarray}
g(\sqrt{s}) & = & g_{0} exp(-\ell \sqrt{s}),
\end{eqnarray}
in order to reproduce these data in the entire range of beam energies. The
parameters, $g_0$, $\lambda$ and $\ell$ were determined by fitting
to the elastic proton-proton and proton-neutron scattering data at the
beam energies in the range of 400 MeV to 4.0 GeV~\cite{sch94,shy96}.
It may be noted that this procedure fixes also the signs of the
effective Lagrangians [Eqs.~(1)-(5)].
The values of various parameters are given in Table 1 of
Ref.~\cite{shy99}. The same parameters for these vertices were also
used in calculations of the pion and the dilepton production in  $pp$
collisions.  Thus we ensure that the $NN$ 
elastic scattering channel remains the same in the description of various
inelastic channels within this approach, as it should be. 
  
Below 2 GeV center of mass (c.m.) energy, only three resonances,
$N^*$(1650), $N^*$(1710), and $N^*$(1720), have significant decay
branching ratios into $KY$ channels. Therefore, we
have considered only these three resonances in our calculations.
The $N^*$(1700) resonance having very small (and uncertain) branching
ratio for the decay to these channels, has been excluded. 
Since all the three resonances can couple to the meson-nucleon channel 
considered in the previous section, we require the effective Lagrangians
for all the four resonance-nucleon-meson vertices corresponding to
all the included resonances. Since the mass of the strange quark is
much higher than that of the $u-$ or $d-$ quark, one does not expect
the pion like strict chiral constraints for the case of other
pseudoscalar mesons like $\eta$ and $K$ (to be called $\zeta$
in the following). Thus, one has a choice of psuedoscalar (PS) or PV
couplings for the $NN\zeta$ and $N^*_{1/2}N\zeta$ vertices (forms of the
corresponding effective Lagrangians are given in Ref.~\cite{shy99}).
The same holds also for the $N^*_{1/2}Y K$ vertices.

In principle, one can select a linear combination of both and fit the
PS/PV ratio to the data. However, to minimize the number of parameters
we choose either PS or PV coupling at a time. In the results shown below,
we have used PS couplings for both $N^*N\pi$ and $N^*\Lambda K^+$
vertices involving spin-1/2 resonances of even and odd parities.
Calculations have also been performed with the corresponding
PV couplings. The cross sections calculated with this option for
the $N_{1/2}^*YK$ vertex deviate very little from those
obtained with the corresponding PS couplings. However, data shows
a clear preference for the PS coupling at the $N_{1/2}^*N\pi$ vertices.

The couplings constants for the vertices involving resonances
are determined from the experimentally observed quantities such as
branching ratios for their decays to corresponding
channels. It may however, be noted that such a procedure can not be
used to determine the coupling constant for the $N^*(1650)\Sigma K$
vertices, as the on-shell decays of this resonance to 
$\Sigma K$ channel are inhibited. Therefore, we have tried to determine this
coupling constant by fitting the available data on the
$\pi^+p \to \Sigma^+ K^+$, $\pi^-p \to \Sigma^0 K^0$, and
$\pi^-p \to \Sigma^-K^+$ reactions in an effective Lagrangian  coupled
channels approach \cite{feu98,pen02}, where all the available data for the
transitions from $\pi N$ to five meson-baryon final states, $\pi N$,
$\pi \pi N$, $\eta N$, $K\Lambda$, and $K\Sigma$ are simultaneously
analyzed for center of mass energies ranging from threshold to 2 GeV. In
this analysis all the baryonic resonances with spin $\leq \frac{3}{2}$
up to excitation energies of 2 GeV are included as intermediate states. 
Since the resonances considered in this study have no known
branching ratios for the decay into the $N\omega$ channel, we determine the
coupling constants for the $N^*N\omega$ vertices by the strict
vector meson dominance (VMD) hypothesis~\cite{saku69}, which  
is based essentially on the assumption that the coupling of photons on hadrons 
takes place through a vector meson.  

It should be stressed that the branching ratios determine only
the square of the corresponding coupling constants; thus their signs remain
uncertain in this method. Predictions from independent calculations
(${\it e.g}$ the quark model) can, however, be used to constrain these
signs. The magnitude as well as signs of the coupling constants for the
$N^*N\pi$, $N^*\Lambda K$, $N^*N\rho$, and $N^*N(\pi \pi)_{s-wave}$
vertices were determined by Feuster and Mosel~\cite{feu98} and Manley
and Saleski~\cite{manl92} in their analysis of the pion-nucleon data
involving the final states $\pi N$, $\pi \pi N$, $\eta N$, and $K\Lambda$.
Predictions for some of these quantities are also given in the
constituent quark model calculations of Capstick and Roberts~\cite{caps94}.
Guided by the results of these studies, we have chosen the positive
sign for the coupling constants for these vertices. Unfortunately,
the quark model calculations for the $N^*N\omega$ vertices are still
sparse and an unambiguous prediction for the signs of the corresponding
coupling constants may not be possible at this stage~\cite{stan93}.
Nevertheless, we have chosen a positive sign for the coupling constants
for these vertices as well.

The resonance properties used in the calculations of the decay widths are
given in Table 1, where the resulting coupling constants and the adopted
values of the cut-off parameters ($\lambda_i^{NN^*}$) are also shown.
It may be noted that we have fixed the latter to one value in order to
minimize the number of free parameters.
\begin{table}[here]
\begin{center}
\caption {Coupling constants and cut-off parameters for the
$N^*N$-meson and $N^*$-hyperon-meson vertices used in the calculations}
\vspace{0.5cm}
\begin{tabular}{|c|c|c|c|c|}
\hline
Resonance  & Decay channel & Branching & $g^2/4\pi$ & cut-off \\
           &               & ratio (GeV)&   & (\footnotesize{GeV}) \\
\hline
$N^*$(1710)& $N\pi$    & 0.150     & 0.0863 & 850.0 \\
           & $N\rho$   & 0.150     & 1.3653 & 850.0 \\
           & $N\omega$ &           & 0.1189 & 850.0 \\
           & $N\sigma$ & 0.170     & 0.0361 & 850.0 \\
           & $\Lambda K$ & 0.150   & 2.9761 &       \\
           & $\Sigma K$  &         &  4.4044&       \\  
$N^*$(1720)& $N\pi$    & 0.100     & 0.0023 & 850.0 \\
           & $N\rho$   & 0.700     & 90.637 & 850.0 \\
           & $N\omega$ &           & 22.810 & 850.0 \\
           & $N\sigma$ & 0.120     & 0.1926 & 850.0 \\
           & $\Lambda K$ & 0.080   & 0.0817 &          \\
           & $\Sigma K$ &          & 0.2204 &       \\  
$N^*$(1650)& $N\pi$     & 0.700     & 0.0521 & 850.0 \\
           & $N\rho$    & 0.08      & 0.5447 & 850.0 \\
           & $N\omega$  &           & 0.2582 & 850.0 \\
           & $N\sigma$  & 0.025     & 0.2882 & 850.0 \\
           & $\Lambda K$ & 0.070    & 0.0485 &       \\
           & $\Sigma K$ &          &  0.0165 &       \\  
\hline 
\end{tabular}
\end{center}
\end{table}

After having established the effective Lagrangians,
coupling constants and form of the propagators (which are given in 
Ref.~\cite{shy99}), it is straight forward to 
write down the amplitudes for various diagrams associated with
the $pp \to pYK$ reactions which can be calculated numerically by following
${\it e.g.}$ the techniques discussed in~\cite{shy96}. The isospin
part is treated separately. This
gives rise to a constant factor for each graph, which is unity for the
reaction under study. It should be noted that the signs of
various amplitudes are fixed by those of the effective Lagrangian
densities, coupling constants and propagators
as described above. These signs are not allowed to change anywhere in
the calculations.

In the present form of our effective Lagrangian theory, the 
energy dependence of the cross section due to FSI is separated from
that of the primary production amplitude and the total amplitude is
written as,
\begin{eqnarray}
A_{fi} & = & M_{fi}(pp \rightarrow pYK^+) \cdot T_{ff},
\end{eqnarray}
where $M_{fi}(pp \rightarrow pYK^+)$ is the primary associated 
$YK$ production amplitude, while $T_{ff}$ describes the re-scattering
among the final particles which goes to unity in the limit of no FSI.
The latter is taken to be the coherent sum of the two-body on-mass-shell
elastic scattering amplitudes $t_i$ (with $i$ going from 1 to 3),
of the interacting particle pairs $j-k$ in the final
channel. This type of approach has been used earlier to
describe the pion~\cite{shy98,dub86,mei98}, $\eta$-meson
\cite{mol96,dru97,del04}, $\Lambda K^+$ \cite{shy99} and $\phi$-meson
\cite{tit00} production in $pp$ collisions.

An assumption inherent in Eq.~(9) is that the reaction takes place over
a small region of space (which is fulfilled rather well in
near threshold reactions involving heavy mesons). Under this condition the
amplitudes $t_i$ can be expressed in terms of the inverse of the
Jost function~\cite{wat52,shy98} which has been calculated by
using a Coulomb modified effective range expansion of the
phase-shift~\cite{noy72}. The required effective range and
scattering length parameters are given in Refs.~\cite{shy99,shy01}.  

\section{Results and Discussions}

The total cross sections for the $\Lambda K^+$ and $\Sigma^0 K^+$
reactions as a function of the excess energy are shown in Fig.~2.
The calculations are the coherent sum of all resonance excitation and meson
exchange processes as described earlier. In both cases a good agreement is 
obtained  between theory and the data available from the COSY-11
collaboration. Keeping in mind the fact that all parameters of the
model, except for those of $N^*Yp$ vertices and the FSI, were the
same in the two calculations and that no parameter was freely varied,
this agreement is quite satisfactory. It should be noted that we do not
require to introduce arbitrary normalization constants to get the agreement
between calculations and the data.  We also show in this figure the
results obtained without including the FSI effects (dashed line). It
can be seen that the FSI effects are vital for a proper description of 
the experimental data in both the cases.
\begin{figure}[H]
\parbox{0.4\textwidth}
   {\epsfig{file=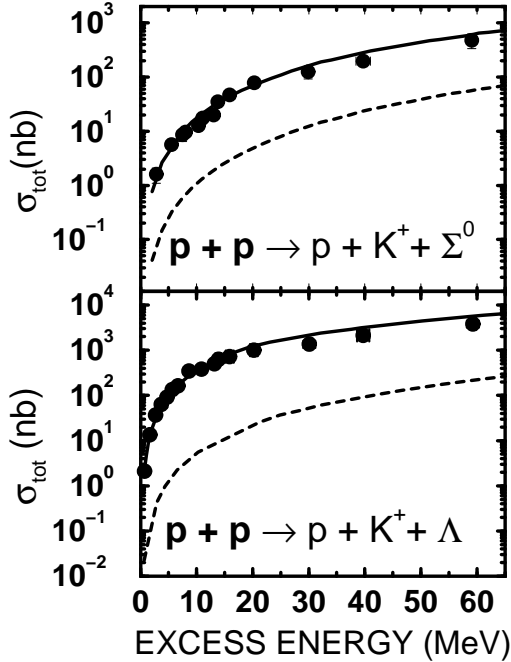,width=0.9\textwidth}}
\hfill
\parbox{0.4\textwidth}
  {\caption{\label{fig.2} {\small  
Comparison of the calculated and the experimental total
cross sections for the $pp \to p\Lambda K^+$ 
and $pp \to p\Sigma^0K^+$ reactions as a function of
the excess energy. Results obtained with no FSI effects
are shown by dashed lines. The experimental data are from
Refs.~\protect\cite{sew99}
  }.}}
\end{figure}
\noindent
In Fig.~3, we have investigated the role of various meson exchange
processes in describing the total cross sections. The dashed, long-dashed,
dashed-dotted, and solid with black square curves represent the
contributions of $\pi$, $\rho$, $\omega$ and $\sigma$ meson exchanges,
respectively. The contribution of the heavy axial meson exchange
is not shown in this figure as it is negligibly small. The coherent
sum of all the meson-exchange processes is shown by the solid line.
It is clear that the pion exchange graphs dominate the production
process for both the reactions in the entire range of beam energies.
Contributions of $\rho$ and $\omega$ meson exchanges are almost
insignificant. On the other hand, the $\sigma$ meson exchange, which
models the correlated $s-$wave two-pion exchange process 
and provides about 2/3 of this exchange in the low energy $NN$
interaction, plays a relatively more important role. This observation
has also been made in case of the $NN\rightarrow NN\pi$
reaction~\cite{dmit86,risk93,horo94, shy96}. 

The individual contributions of various nucleon resonances to the
total cross sections of the two reactions are shown in Fig.~4. 
We note that in both
the cases, the cross section is dominated by the $N^*$(1650)
resonance excitation for $\epsilon < 30$ MeV. Since $N^*$(1650) is
the lowest energy baryonic resonance having branching
ratios for the decay to $YK^+$ channels, its dominance in
\begin{figure}[H]
\parbox{0.4\textwidth}
   {\epsfig{file=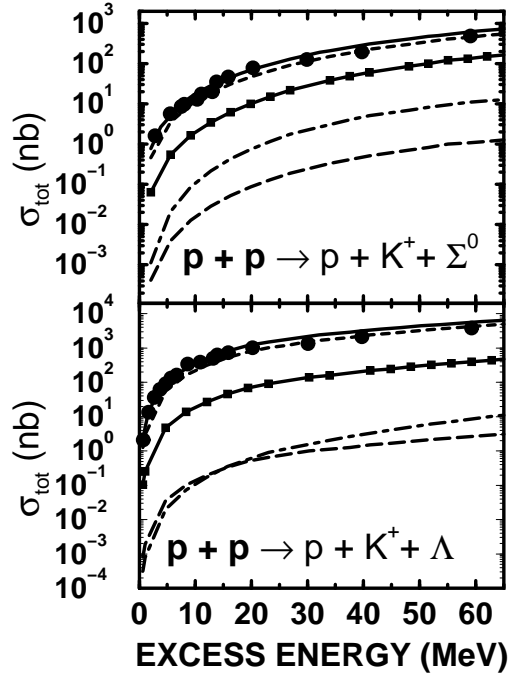,width=0.9\textwidth}}
\hfill
\parbox{0.4\textwidth}
  {\caption{\label{fig.3} {\small 
Contributions of various meson exchange processes to the
total cross section for the $pp \rightarrow pK^{+}\Lambda$
and $pp \rightarrow pK^{+}\Sigma^0$ reactions. The dashed, long-dashed,
dashed-dotted and solid with black squares curves represent the
contributions of $\pi$, $\rho$, $\omega$ and $\sigma$ meson exchanges,
respectively. Their coherent sums are shown by the solid lines }.}}
\end{figure}
\noindent
these
reactions at beam energies very close the kaon production threshold
is to be expected. In the near threshold region the relative dominance
of various resonances is determined by the dynamics of the reaction
where the difference of about 60 MeV in excitation energies of
$N^*$(1650) and $N^*$(1710) resonances plays a crucial role.
 
However, for $\epsilon$ values between 30 - 60 MeV, while the
$pp \to pK^+\Lambda$ reaction continues to be dominated by $N^*(1650)$
excitation, the $pp \to pK^+\Sigma^0$ reaction gets significant
contributions also from $N^*(1710)$ and $N^*(1720)$ resonances. This
difference in the role of the three resonances in the two cases can be
understood in the following way. For a resonances to contribute
significantly, we should have  $m_Y + m_K + \epsilon \geq m_R +
\Gamma_R/2$, where $m_R$ and $\Gamma_R$ are the mass and width of the
resonance, respectively. Therefore, in the region of excess energies
$\geq Q [ = (m_R + \Gamma_R/2) - (m_Y + m_K)]$, the particular resonance
$R$ should contribute significantly. The values of $Q$ for the $pp
\to pK^+\Lambda$ reaction, are 115 MeV, 150 MeV, and 185 MeV, for
the $N^*(1650)$, $N^*(1710)$, and $N^*(1720)$ resonances, respectively.
On the other hand, for the $pp \to p K^+\Sigma^0$ case, they are 36 MeV,
72 MeV and 105 MeV, respectively for these three resonances. Therefore,
contributions of the $N^*(1710)$ and $N^*(1720)$ resonance excitations are
negligibly small for the $K^+\Lambda$ production in the entire energy
range of the COSY-11 data (i.e., for $\epsilon \leq$ 60. MeV) while they are
significant for the $K^+\Sigma^0$ case for $\epsilon >$ 30 MeV. It would
be helpful to have data on the invariant mass spectrum at these excess
energies in order to  confirm these theoretical observations.
\vskip -1.0in
\begin{figure}[H]
\epsfig{file=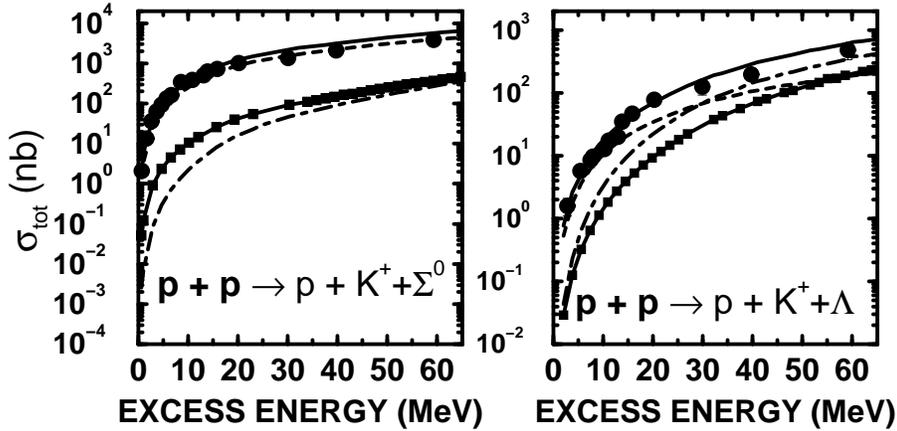,width=1.0\textwidth}
\vskip -0.4in
\caption{\label{fig.4} {\small 
Contributions of N$^{*}$(1650) (dashed line), N$^{*}$(1710)
(full line with black squares) and N$^{*}$(1720) (dashed-dotted line)
baryonic resonances to the total cross section for the two reactions
studied in Fig.~2. Their coherent sum is shown by the solid line.}}
\end{figure}
\noindent
In Fig.~5, we compare our calculations with the data
for the ratio $R$ as a function of $\epsilon$. We have shown here the
results for excess energies up to 60 MeV, where the COSY-11 data are 
available. It is clear that our calculations are able to describe well 
the trend of the fall-off of $R$ as a function of the excess energy. 
It should be noted that  FSI effects account for about $60\% - 80\%$
of the observed ratio for $\epsilon < 30$ MeV and about 
50$\%$ of it beyond this energy. Therefore, not all of the observed value of
$R$ at these beam energies can be accounted for by the FSI effects, which
is in agreement with the observation made in \cite {gas00}. It should
again be emphasized that without considering the contributions of the
$N^*$(1650) resonance for the $\Sigma^0K^+$ reactions, the
calculated ratio would be at least an order of magnitude larger.
Therefore, these data are indeed sensitive to the details of the
reaction mechanism. At higher beam energies ($\epsilon$ $>$ 300 MeV),
values of $R$ obtained with and without FSI effects are almost identical.
In this region the reaction mechanism is different; here all the three
resonances contribute in one way or the other, their
interference terms are significant~\cite{shy99}, and FSI related
effects are unimportant.
\vskip -0.3in
\begin{figure}[H]
\epsfig{file=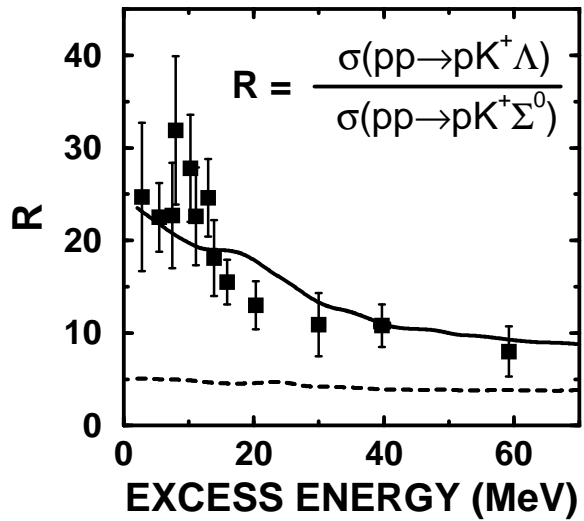,width=1.0\textwidth}
\vskip -0.4in
\caption{\label{fig.5} {\small
Ratio of the total cross sections for
$pp \to p\Lambda K^+$ and $pp \to p\Sigma^0 K^+$ reaction as a function
of the excess energy. The solid and dashed lines show the results of our
calculations with and without FSI effects, respectively. The data are from
\cite{sew99}. }}
\end{figure}
\noindent

\section{Summary and conclusions}

In summary, we have studied the $pp \to p\Lambda K^+$ and
$pp \to p\Sigma^0K^+$ reactions within an effective Lagrangian model.
Most of the parameters of the model are fixed by fitting to the elastic
$NN$ T-matrix which restricts the freedom of varying them freely in
order to fit the data. The reactions proceed via  excitation of the 
$N^*$(1650), $N^*$(1710), and $N^*$(1720) intermediate baryonic resonant
states. An important result of our study is that the $N^*$(1650) resonant
state contributes predominantly to both these reactions at near threshold
beam energies. Therefore, these reactions in this energy regime, provide
an ideal means of investigating the properties of this $S_{11}$ baryonic
resonance. To the extent that the final state interaction effects in the
exit channel can be accounted for by the Watson-Migdal theory, our model
is able to explain the experimentally observed large ratio of the total
cross sections of the two reactions in the near threshold region. 

This work is supported by the Wenner-Gren Center Foundation, Stockholm.
The author wishes to express his sincere thanks to Anders Ingemarsson
and Bo H\"oistad for their very kind hospitality in the Department
of Radiation Science of the Uppsala University. He also wishes to 
acknowledge useful discussions with Bo H\"oistad, G\"oran F\"aldt, 
Ulrich Mosel, and Walter Oelert.

\end{document}